\documentclass[
fleqn,
aps,prb,
superscriptaddress,
preprint,
nobibnotes, 
nofootinbib,
]{revtex4}

\usepackage[]{amsmath}
\usepackage{amssymb}
\usepackage[dvips]{graphicx}
\usepackage{color}
\usepackage{tabularx}
\usepackage{mathtools}
\usepackage{algpseudocode}
\usepackage{enumitem}
\usepackage{multirow}
\usepackage{epstopdf}  
\usepackage{bm}
\usepackage{longtable}
\usepackage{booktabs}

\usepackage{makecell}
\usepackage{hyperref}
\usepackage[version=3]{mhchem}
\usepackage{tikz}
\usetikzlibrary{shapes,snakes}
\usetikzlibrary{arrows}
\usepackage{threeparttable}

\usepackage{xparse}
\usepackage[capitalize,noabbrev]{cleveref}

\newcommand{\hmn}[1]{
  \ensuremath{\begingroup\setupHMN #1\endgroup}%
}
\newcommand{\setupHMN}{%
  \doHMN{-}{\HMNoverline}%
  \doHMN{*}{\HMNminverse}%
  \doHMN{i}{\infty}
  \doHMN{_}{\HMNsubscript}
}
\newcommand{\doHMN}[2]{%
  \begingroup\lccode`~=`#1
  \lowercase{\endgroup\let~}#2%
  \mathcode`#1="8000
}
\newcommand{\HMNminverse}[1]{\frac{#1}{m}}

\newcommand{\HMNoverline}[1]{\mkern1mu\overline{\mkern-1mu#1\mkern-1mu}\mkern1mu}
\newcommand{\HMNsubscript}[1]{_{#1}}

\newcommand{\scha}{\mathrm{SCHA}}

\begin{document}
\title{
Crystal structure prediction with nuclear quantum and finite-temperature effects via deep free energy learning
}

\author{Xiaoyang Wang}
\affiliation{National Key Laboratory of Computational Physics,
  Institute of Applied Physics and Computational Mathematics, Fenghao East Road 2, Beijing 100094, P.R.~China}

\author{Yinan Wang}
\affiliation{Artificial Intelligence for Science Institute, No.150 Chengfu Road, Haidian District, Beijing, 100084, China}

\author{Wenbo Zhao}
\affiliation{Key Laboratory of Material Simulation Methods and Software of Ministry of Education, College of Physics, Jilin University, Changchun 130012, China}
\affiliation{International Center of Future Science, Jilin University, Changchun 130012, China} 

\author{Hanyu Liu}
\affiliation{Key Laboratory of Material Simulation Methods and Software of Ministry of Education, College of Physics, Jilin University, Changchun 130012, China}
\affiliation{International Center of Future Science, Jilin University, Changchun 130012, China}

\author{Hao Xie}
\affiliation{Institute of Physics, École Polytechnique Fédérale de Lausanne (EPFL), CH-1015 Lausanne, Switzerland}

\author{Lei Wang}
\affiliation{Institute of Physics, 
Chinese Academy of Sciences, Beijing 100190, China}

\author{Han Wang}
\thanks{Corresponding Author Han Wang, wang\textunderscore han@iapcm.ac.cn}
\affiliation{National Key Laboratory of Computational Physics,
  Institute of Applied Physics and Computational Mathematics, Fenghao East Road 2, Beijing 100094, P.R.~China}
\affiliation{HEDPS, CAPT, College of Engineering, Peking University, Beijing 100871, P.R.~China}

\begin{abstract}
Accurate crystal structure prediction (CSP) requires accounting for finite-temperature and nuclear quantum effects, yet first-principles evaluation of the free energy surface (FES) remains prohibitive for high-throughput searches.
We observe that the self-consistent harmonic approximation (SCHA) FES, as a function of nuclear centroid positions, shares the same mathematical structure as a potential-energy surface and can therefore be directly learned by a deep neural network potential.
The resulting deep free energy (DF) model, constructed via a two-level concurrent-learning workflow, evaluates free energies, forces, and stresses in a single forward pass.
Applied to the La--Sc--H system at 200~GPa and 300~K, DF-based CSP reproduces the stability of the experimentally observed \ce{LaH10} and \ce{LaSc2H24}, and discovers an unreported thermodynamically stable clathrate hydride: \hmn{P4/mmm} \ce{LaScH8}.
Benchmarked on the \ce{LaH10} system, the DF model achieves a $1.72\times 10^6$-fold cost reduction relative to DFT-level SSCHA.
The DF framework provides a scalable route for incorporating finite-temperature and nuclear quantum effects into high-throughput crystal structure prediction.
\end{abstract}

\maketitle


\section{Introduction}

Crystal structure prediction (CSP) has become an indispensable tool for computational materials discovery, providing promising candidates for experimental synthesis and substantially reducing the trial-and-error effort inherent in traditional approaches~\cite{Wang2015Materials}.
CSP methods have facilitated breakthroughs across diverse domains, including high-temperature superconductivity~\cite{Sun2024NSRClathrate}, thermoelectrics~\cite{nunez2018efficient}, superhard materials~\cite{kvashnin2017computational}, and geoscience~\cite{hermann2019geoscience}.

The primary goal of CSP is to identify the global minimum of the high-dimensional classical potential-energy surface (PES) that defines the zero-temperature stability of a crystal, without accounting for nuclear quantum effects.
A variety of global optimization strategies have been developed for this purpose, including simulated annealing~\cite{kirkpatrick1983optimization}, evolutionary algorithms~\cite{lonie2011xtalopt}, basin hopping~\cite{wales1997global}, random sampling~\cite{pickard2011ab}, and particle swarm optimization~\cite{eberhart1995particle}, and have been implemented in widely used packages such as AIRSS~\cite{liborio2018computational}, USPEX~\cite{glass2006uspex}, MAGUS~\cite{wang2023magus}, and CALYPSO~\cite{wang2012calypso}.
Combined with first-principles calculations, these methods have made CSP a routine tool for computational materials discovery.

A key gap between current CSP methods and real-world materials discovery is that predictions are based on the classical PES, which does not capture finite-temperature effects (FTE) or nuclear quantum effects (NQE) that may significantly contribute to the stability of crystal structures.
For instance, $\beta$-Ti, a high-temperature body-centered-cubic (BCC) structure~\cite{van2006experimental}, is found to be dynamically unstable in the BCC configuration when optimized using standard density functional theory (DFT) methods~\cite{souvatzis2008entropy}.
Another example is the ternary hydride \hmn{P6/mmm}  \ce{LaSc2H24}, recently demonstrated experimentally to exhibit room-temperature stability and superconductivity with an onset temperature of 271-298 kelvin at 195-266 GPa~\cite{LaSc2H24_syn}.
However, this phase is dynamically unstable on the classical PES at 200 GPa.
These examples demonstrate that CSP relying solely on the classical PES would miss important structures stabilized by FTE and NQE.

Modeling crystal stability with FTE and NQE requires the construction of the free-energy surface (FES), and the corresponding CSP task reduces to identifying the local minima on the FES.
The precise computation of the FES is nontrivial, as it requires integration of the partition function over the high-dimensional configuration space.
The simplest approximation to the FES is provided by the harmonic approximation, which expands the crystal potential energy to second order in atomic displacements~\cite{born1996dynamical}.
To go beyond this limit, several systematic extensions have been developed, including the quasi-harmonic approximation, which incorporates volume-dependent free-energy contributions~\cite{dove1993introduction}, and perturbative anharmonic approaches that expand the PES to higher orders beyond the quadratic term~\cite{cowley1963lattice}.
Alternatively, the FES can be estimated from molecular dynamics (MD) simulations, either by extracting temperature-dependent effective force constants~\cite{hellman2011lattice} or via thermodynamic integration~\cite{frenkel1984new,vovcadlo2002ab}.
To explicitly incorporate NQE in addition to FTE, classical MD can be replaced by path-integral molecular dynamics (PIMD)~\cite{ceperley1995path}.
Beyond perturbative treatments, non-perturbative frameworks such as the self-consistent harmonic approximation (SCHA)~\cite{hooton1955li} and its modern stochastic formulation, the stochastic self-consistent harmonic approximation (SSCHA)~\cite{errea2014anharmonic,bianco2017second,bianco2018high,monacelli2021stochastic}, provide a variational description of anharmonic and quantum nuclear effects.
Standard SSCHA implementations rely on first-principles methods, typically DFT, to evaluate the potential-energy contribution to the FES.
More recently, these approaches have been significantly accelerated by replacing computationally intensive DFT calculations with machine-learning interatomic potentials (MLIPs)~\cite{belli2025efficient,gao2025iterative,poletaev2025sscha}.

Despite rapid advances in FES methodologies, direct FES evaluation remains computationally demanding, and high-throughput CSP on the FES is still in its early stages.
Many existing approaches conduct structure searches on the PES and subsequently evaluate the finite-temperature stability of selected candidates in a \textit{post hoc} manner using thermodynamic integration~\cite{wang2023data} or solid--liquid phase coexistence molecular dynamics~\cite{wang2024high}.
Kruglov \textit{et al.}~\cite{kruglov2023crystal} proposed the T-USPEX framework, which incorporates free-energy evaluations at each CSP iteration; however, candidate structures are still generated via PES-based searches.
More recently, MLIP-accelerated SSCHA has been integrated directly into evolutionary CSP workflows~\cite{belli2025efficient,gao2025iterative,poletaev2025sscha}, enabling structure relaxation on the anharmonic FES at a fraction of the first-principles cost.
Nevertheless, these approaches still require explicit SSCHA free-energy minimizations, each involving hundreds of MLIP evaluations, for every candidate structure at every CSP iteration, which limits the throughput of the structure search.

In this work, we observe that the SCHA free-energy surface, as a function of centroid positions alone, shares the same mathematical structure as a potential-energy surface, and can therefore be directly learned by a deep neural network potential architecture. We term the resulting model the deep free energy (DF).
The DF model is constructed via a two-level concurrent-learning workflow: the first level generates a deep-learning potential-energy model (DP) as a surrogate for computationally expensive DFT calculations, while the second level constructs a DF model that approximates the SCHA free-energy surface.
Once trained, the DF model evaluates the free energy, free-energy forces, and free-energy stresses in a single forward pass, entirely bypassing the stochastic SSCHA minimization required by existing MLIP-based approaches.
CSP is then performed directly on this DF-approximated FES to identify structures that are stable with respect to free-energy minimization.
The proposed framework is applied to the La--Sc--H ternary system, where a previously unreported thermodynamically stable clathrate hydride \hmn{P4/mmm} \ce{LaScH8} is discovered.

The remainder of this paper is organized as follows.
Section~\ref{sec:method} introduces the self-consistent harmonic approximation (SCHA) for the free-energy surface, its stochastic formulation (SSCHA), and the proposed deep-learning free-energy modeling workflow.
Section~\ref{sec:result} presents the application of the DF method to the La--Sc--H ternary system, demonstrating that the approach not only rediscovers the well-known \hmn{P6/mmm} \ce{LaSc2H24} but also reveals a previously unreported structure.
Conclusions and discussion are provided in Section~\ref{sec:conclude}.

\section{Method}\label{sec:method}

\subsection{Self-consistent harmonic approximation}

The self-consistent harmonic approximation (SCHA) applies to both molecular systems under open boundary conditions and crystalline systems under periodic boundary conditions.
For clarity, the formulation below is presented at the $\Gamma$ point only; extension to multiple Brillouin-zone sampling points is straightforward.
We consider an atomistic system composed of $N$ nuclei, whose quantum Hamiltonian is given by
\begin{equation}
    H = K + V(R),
\end{equation}
where $K$ is the nuclear kinetic energy operator and $V(R)$ is the potential energy surface under the Born--Oppenheimer approximation. Here, $R \in \mathbb{R}^{3N}$ denotes the Cartesian coordinates of all nuclei.
At finite temperature $T$, the Helmholtz free energy of the system is defined as
\begin{equation}\label{eq:fe-def}
    A = \mathrm{Tr}[\rho H] + \frac{1}{\beta}\,\mathrm{Tr}[\rho \ln \rho],
\end{equation}
where $\beta = (k_B T)^{-1}$ and $k_B$ is the Boltzmann constant. The equilibrium density matrix $\rho$ is given by $\rho = e^{-\beta H}/Z$, with the partition function defined as $Z = \mathrm{Tr}[e^{-\beta H}]$.

Within the SCHA, the exact equilibrium density matrix of the physical system is approximated by a trial density matrix,
\begin{align}
    \tilde{\rho} = \frac{1}{\tilde Z} e^{-\beta \tilde H}, 
    \qquad 
    \tilde Z = \mathrm{Tr}\!\left[e^{-\beta \tilde H}\right],
\end{align}
where the trial Hamiltonian $\tilde H$ is chosen to be harmonic,
\begin{align}\label{eq:trial-h}
    \tilde H(\mathcal R, \Phi) = K
    + \frac{1}{2} \sum_{ab} \Phi_{ab}
    \left(R_a - \mathcal R_a\right)
    \left(R_b - \mathcal R_b\right).
\end{align}
Here, $\mathcal R \in \mathbb{R}^{3N}$ and $\Phi \in \mathbb{R}^{3N \times 3N}$ are variational parameters, with $\Phi$ constrained to be positive definite.
Under the trial density matrix $\tilde{\rho}$, the expectation value of the nuclear coordinates satisfies
$\langle R \rangle_{\tilde{\rho}} = \mathcal R$,
so that $\mathcal R$ represents the centroid of the nuclear probability distribution~\cite{bianco2017second}, while $\Phi$ characterizes the quantum and thermal fluctuations around this centroid.
Substituting the physical density matrix in Eq.~\eqref{eq:fe-def} with the trial density matrix and invoking the Gibbs--Bogoliubov variational principle~\cite{isihara1968gibbs}, one obtains
\begin{align}\label{eq:fe-min-gb}
    A[\tilde{\rho}]
    = \mathrm{Tr}[\tilde{\rho} H]
    + \frac{1}{\beta}\,\mathrm{Tr}[\tilde{\rho} \ln \tilde{\rho}]
    \geq A,
\end{align}
where $A$ denotes the exact Helmholtz free energy of the system.
The exact Helmholtz free energy therefore constitutes a lower bound to the variational free energy $A[\tilde{\rho}]$, and the bound is saturated if and only if the trial density matrix coincides with the exact equilibrium density matrix, $\tilde{\rho} = \rho$.

Since the trial density matrix is parameterized by the centroid $\mathcal R$ and the fluctuation matrix $\Phi$, the variational problem in Eq.~\eqref{eq:fe-min-gb} can be reformulated as
\begin{align}\label{eq:fe-min-sscha}
\min_{\tilde{\rho}} A[\tilde{\rho}]
= \min_{\mathcal R,\,\Phi} A(\mathcal R,\Phi)
= \min_{\mathcal R}\Bigl[\min_{\Phi} A(\mathcal R,\Phi)\Bigr].
\end{align}
This motivates the definition of the SCHA free energy,
\begin{align}\label{eq:fe-def-scha}
A_{\scha}(\mathcal R) = \min_{\Phi} A(\mathcal R,\Phi),
\end{align}
such that the original variational problem reduces to
\begin{equation}\label{eq:fe-min-scha}
\min_{\tilde{\rho}} A[\tilde{\rho}] = \min_{\mathcal R} A_{\scha}(\mathcal R).
\end{equation}
Within the SCHA framework, crystal structure prediction is therefore recast as a minimization over centroid positions $\mathcal R$ on the SCHA free energy surface.
This formulation closely parallels conventional CSP, which seeks minima of the potential energy surface $V(R)$.
The key distinction is that $A_{\scha}(\mathcal R)$ explicitly incorporates finite-temperature and nuclear quantum effects, whereas standard CSP optimizes the zero-temperature potential energy.
Owing to this shared mathematical structure, established CSP algorithms, such as CALYPSO, can be directly adapted to perform structure searches on the SCHA free energy surface.
We note that the simulation cell matrix $\mathbf{h}$ enters $A_{\scha}$ on an equal footing with the centroid coordinates $\mathcal R$, but is suppressed in the notation for conciseness.
In the high-pressure structure searches considered in this work, the $PV$ contribution is always included, so that the target function is the Gibbs free energy $G_{\scha}(\mathcal R, \mathbf{h}) = A_{\scha}(\mathcal R, \mathbf{h}) + P\det(\mathbf{h})$.

\subsection{Stochastic self-consistent harmonic approximation}

Under the trial density matrix ansatz, most contributions to the variational free energy in Eq.~\eqref{eq:fe-min-gb} can be evaluated analytically, with the exception of the potential energy term $\mathrm{Tr}[\tilde{\rho} V]$. 
In the modern implementation of the self-consistent harmonic approximation, known as the stochastic self-consistent harmonic approximation (SSCHA), \cite{errea2014anharmonic,monacelli2021stochastic}, this term is estimated stochastically as
\begin{align}\label{eq:sscha-esti}
    \mathrm{Tr}[\tilde{\rho} V] 
    \approx 
    \frac{1}{M_s} \sum_{i=1}^{M_s} V(R_i),
\end{align}
where $\{R_i\}_{i=1}^{M_s}$ are configurations sampled from the trial density matrix $\tilde{\rho}$. 
During the variational minimization, the gradients of the free energy with respect to both the centroid $\mathcal R$ and the fluctuation matrix $\Phi$ are evaluated using analogous stochastic estimators.
Further implementation details of the SSCHA algorithm can be found in Ref.~\cite{errea2014anharmonic}.

The estimator in Eq.~\eqref{eq:sscha-esti} is unbiased and converges to the exact expectation value in the limit $M_s \to \infty$. 
However, this stochastic evaluation substantially increases the computational cost, as each estimation of the free energy and its gradients requires $M_s$ evaluations of the potential energy $V(R)$, with $M_s$ typically ranging from $10^2$ to $10^3$. 
Consequently, free energy minimization within the SSCHA framework is significantly more expensive than conventional potential energy surface minimization, where each optimization step typically involves a single energy evaluation.
This cost becomes prohibitive when first-principles methods, such as density functional theory (DFT), are employed to compute $V(R)$.

\subsection{Deep learning free energy}
\label{sec:method-df}

\begin{figure}
\begin{center}
\includegraphics[width=1.0\textwidth]{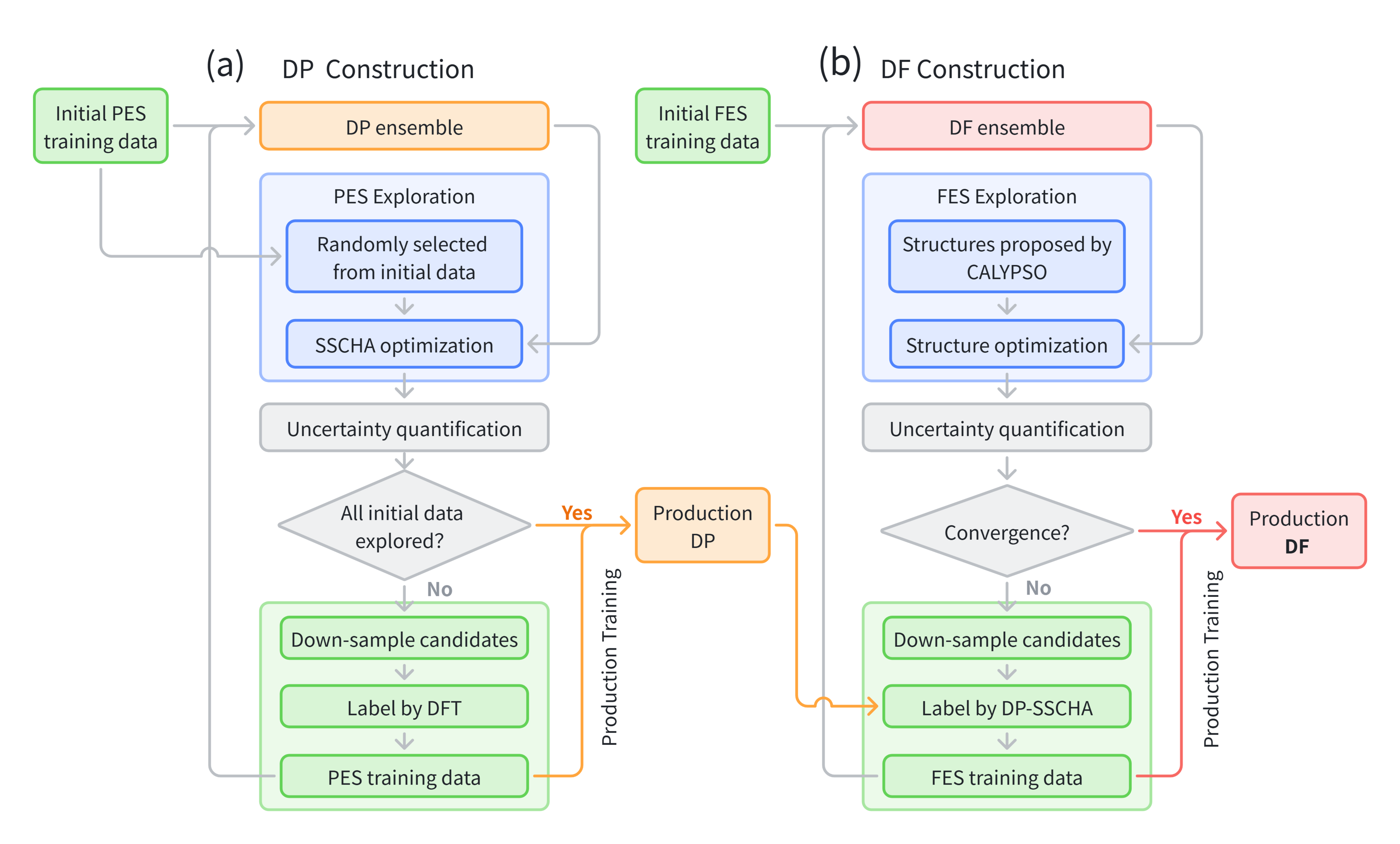}\\[5pt]  
\caption{Two-stage concurrent-learning workflow for constructing the deep free energy (DF) model. (a)~First stage: a DP model is iteratively trained on DFT-labeled configurations sampled using SSCHA minimizations, leading to a production DP that replaces DFT. (b)~Second stage: a DF model is iteratively trained on DP-SSCHA-labeled structures generated by CSP on the free-energy surface, converging to a production DF that directly predicts free energies, forces, and stresses.} \label{fig:workflow}
\end{center}
\end{figure}

We propose a deep-learning framework, termed deep free energy (DF), that directly approximates the SCHA free-energy surface $A_{\scha}(\mathcal R)$, thereby enabling efficient structure searches using established CSP algorithms.
The ultimate target is the free-energy surface obtained by performing SSCHA with the DFT Born--Oppenheimer potential energy, hereafter referred to as the DFT-SSCHA free-energy surface.
This target is reached through two successive levels of approximation (Fig.~\ref{fig:workflow}): first, a deep-learning potential (DP) model replaces DFT as the potential-energy evaluator within the SSCHA, yielding the DP-SSCHA free-energy surface; second, the DP-SSCHA free-energy surface is learned by the DF model.
In Sec.~\ref{sec:acc-models}, we show that the errors introduced at each level are well controlled and that the overall DF-versus-DFT-SSCHA discrepancy remains small.

The first stage constructs a DP model to replace DFT in the SSCHA potential-energy evaluation.
The framework is agnostic to the choice of DP architecture~\cite{poletaev2025sscha,belli2025efficient}. We utilize DPA1\cite{DPA1} for dataset construction and adopt DPA3\cite{zhang2025graph} for the production DP model in this work.
We start from an initial PES training dataset generated for zero-temperature CSP tasks. Such a dataset is generally insufficient for SCHA-based free-energy evaluation, because configurations sampled from the SCHA trial density matrix can deviate substantially from those encountered along zero-temperature relaxation trajectories, particularly when anharmonicity or quantum fluctuations are pronounced.

To bridge this distributional mismatch, the framework employs an iterative concurrent-learning strategy~\cite{zhang2019active,zhang2020dp} to systematically enrich the training dataset (Fig.~\ref{fig:workflow}(a)).
Starting from the initial dataset, an ensemble of four DP models is trained using identical network architectures and hyperparameters, differing only in the random initialization of their parameters.
This model ensemble is subsequently used both to explore the configuration space and to quantify uncertainties in atomic force predictions via model deviation.

Candidate crystal structures for improving the generalizability of the DP within the SSCHA context are generated through a dedicated PES exploration step.
In this step, structures are randomly drawn from the initial CSP-task PES training dataset, their atomic coordinates are taken as the centroids of the nuclear distributions, and SSCHA free-energy minimizations (Eq.~\eqref{eq:fe-min-scha}) are carried out at 300~K and 200~GPa.
Throughout the SSCHA minimization, the DP is used to evaluate the potential energy and atomic forces, yielding substantial computational savings relative to first-principles DFT calculations.
All atomic configurations sampled during the stochastic estimation of the averaged potential energy in Eq.~\eqref{eq:sscha-esti} are recorded for subsequent uncertainty quantification. 

Model uncertainty for the sampled configurations is quantified by the maximum standard deviation of atomic forces predicted across the ensemble of models.
An upper threshold $\sigma_{\mathrm{hi}}$ is introduced to discard configurations with excessively large force uncertainties, which typically correspond to unphysical structures (e.g., unrealistically short interatomic distances).
The remaining configurations are ranked in descending order of force deviation, and the top 10\%, representing the most informative samples for model improvement, are selected for subsequent labeling.
The selected candidate configurations are further sub-sampled to a fixed number and labeled with first-principles DFT calculations of energies, forces, and stresses.
These newly labeled configurations are then appended to the PES training dataset, a new ensemble of DP models is trained on the enriched dataset, and the next concurrent-learning iteration is initiated.

As the iterations proceed, an increasing number of configurations representative of the SSCHA density matrix are incorporated into the training set, progressively improving the accuracy of the DP for SSCHA calculations.
The iterative procedure continues until all structures in the initial dataset have been explored.
Then, a final production-quality DP model based on DPA3 is trained on the complete aggregated dataset with an extended training schedule to achieve higher accuracy than the ensemble models employed during the iterative stage.
The details of the PES construction are presented in Supplemental Information Sec.~I.

The second stage constructs the DF model itself.
Owing to the close mathematical analogy between the SCHA free-energy surface and the potential-energy surface, the same DPA3 architecture can be directly adapted for free-energy learning.
This stage likewise employs an iterative concurrent-learning strategy, closely mirroring the first-stage procedure but with the learning target replaced by the SCHA free-energy surface (Eq.~\eqref{eq:fe-def-scha}).
Accordingly, the two stages differ in how structures are explored and how labels are generated.
In the DF concurrent learning, crystal structures generated by the CSP algorithm (CALYPSO in this work) are interpreted as centroids of the trial density matrix and relaxed directly on the SCHA free-energy surface represented by the current DF model ensemble.
The labels, SCHA free energies together with their gradients with respect to the centroid coordinates (free-energy forces) and the simulation cell (free-energy stresses), are computed via the SSCHA using the previously trained DP model.
The initial FES training dataset is prepared in the same manner: candidate structures are randomly generated by CALYPSO, and their DP-SSCHA free energies, free-energy forces, and free-energy stresses are computed as labels.

Both the DF models used during the concurrent-learning iterations and the final production DF models are trained using a multi-task scheme~\cite{zhang2024dpa}.
Specifically, the model architecture consists of a shared backbone with two prediction heads: one for PES prediction and one for FES prediction. The backbone and both heads are trained jointly end to end.
This multi-task training strategy significantly improves the generalizability of the DF model by leveraging the chemical and configurational priors encoded in the PES data, as will be demonstrated in Sec.~\ref{sec:result}.
The details of the DF model training and FES dataset construction are presented in Supplemental Information Sec.~II.

\section{Results} \label{sec:result}

\subsection{Accuracy of Models}\label{sec:acc-models}

The proposed DF method is used to construct the FES for the La--Sc--H ternary hydride system at $P = 200$~GPa and $T = 300$~K.
The DP model is built to cover a compositional range of $\text{La}_m\text{Sc}_n\text{H}_x$ with $m \in [0,\, 4]$ and $n \in [0,\, 4]$, whereas the DF model covers $m \in [1,\, 4]$ and $n \in [1,\, 4]$, with both models spanning $x \in [2(m+n),\, 10(m+n)]$.
In total, 218,153 DFT-labeled data points (configurations with corresponding potential energies, atomic forces, and virial tensors) are generated for training the DP model, while 5,962 DP-SSCHA data points are generated for multi-task training of the DF model.
The DP training data are split into training and validation sets in a 11:1 ratio, and the DF training data are split similarly in a 9:1 ratio.

\begin{table}
  \scriptsize
  \begin{threeparttable}

  \caption{Training and validation mean absolute errors (MAEs) of the DP and DF models. DF (MT) and DF (ST) denote the multi-task and single-task DF models, respectively. The units for energy ($E$), free energy ($A$), atomic forces ($F$), and virial tensor ($\Xi$) are meV/atom, meV/atom, meV/\AA{}, and meV/atom, respectively.}
  \label{table1}
  {
  \setlength{\tabcolsep}{2.5 mm}{
  \begin{tabular}{c|ccc|ccc|ccc}
\toprule
\hline\hline
Model & \multicolumn{3}{c|}{DP} & \multicolumn{3}{c|}{DF (MT)}&\multicolumn{3}{c}{DF (ST)} \\ \hline
MAE &  $E$ & $F$ & $\Xi$  & $A$ & $F$& $\Xi$& $A$ & $F$& $\Xi$ \\ \hline
Training set  & 9.21 & 134 &36.4 & 6.30&35.7&12.2&6.38&31.9&11.3\\
Validation set & 10.2 & 134 &39.3 &16.7&88.1& 23.8&18.8&98.5&28.0\\ 
\hline\hline
\bottomrule
  \end{tabular}
}
}
  \end{threeparttable}
\end{table}

We first evaluate the accuracy of the DF and DP models against their respective learning targets, DP-SSCHA and DFT (Table~\ref{table1}).
The DP model yields mean absolute errors (MAEs) of $9.21$~meV/atom for energy ($E$), $134$~meV/\AA{} for atomic forces ($F$), and $36.4$~meV/atom for virial ($\Xi$) on the training set.
On the validation set, it achieves $10.2$~meV/atom, $134$~meV/\AA{}, and $39.3$~meV/atom, respectively, demonstrating robust generalizability.
The DF model exhibits higher overall accuracy than the DP model on the training set, yet a noticeable gap exists between its training and validation MAEs.
Nevertheless, the validation errors of the DF (MT) model relative to the DP-SSCHA target, namely $16.7$~meV/atom for free energy ($A$), $88.1$~meV/\AA{} for force ($F$), and $23.8$~meV/atom for virial ($\Xi$)~\footnote{In the context of free energy, force and virial should be understood as the derivatives of the free energy with respect to the atomic coordinates and the simulation cell, respectively.}, remain comparable to those of the DP model relative to DFT.
This gap may be attributed to the stochastic nature of free-energy surface evaluation in the DP-SSCHA method.
Comparing the two DF variants, the DF (ST) model achieves slightly lower training errors than DF (MT), yet its validation errors are consistently higher, indicating overfitting to the training data.
This confirms that the multi-task training strategy improves the generalizability of the DF model by sharing representations with the PES prediction task.

\begin{figure}
\begin{center}
\includegraphics[width=.5\textwidth]{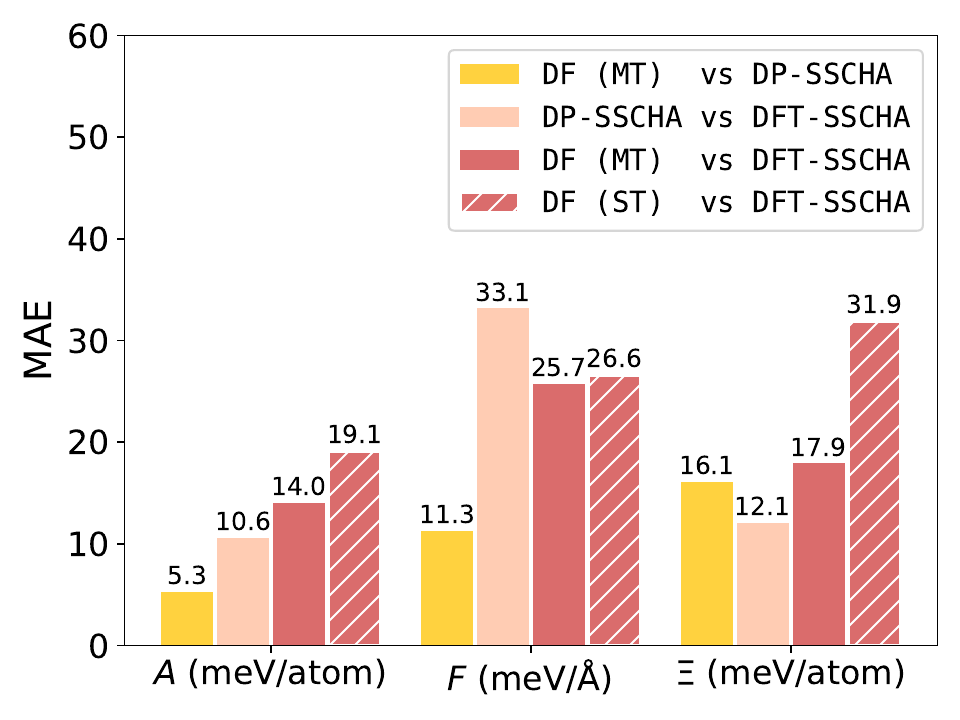}\\[5pt]  
\caption{Mean absolute errors (MAEs) of the free energy ($A$), forces ($F$), and virial ($\Xi$) evaluated on seven representative binary and ternary configurations at 300~K. Results are shown for the DF multi-task (MT) and single-task (ST) models, and the DP-SSCHA method, each compared against DFT-SSCHA as the reference. The decomposition of the DF (MT) error into its two levels of approximation, DF (MT) vs DP-SSCHA and DP-SSCHA vs DFT-SSCHA, is also shown.} \label{fig:fig2}
\end{center}
\end{figure}

The overall accuracy of the DF (MT) model is assessed against the DFT-SSCHA free-energy surface on a test set comprising seven representative binary and ternary configurations at 300~K: \hmn{P4/mmm} \ce{LaH4}, \hmn{Fm-3m} \ce{LaH10}, \hmn{Fm-3m} \ce{ScH3}, \hmn{P6_3mmc} \ce{ScH6}, \hmn{Im-3m} \ce{ScH6}, \hmn{P4/mmm} \ce{LaScH8}, and \hmn{P6/mmm} \ce{LaSc2H24}.
DFT-SSCHA free energies are computed for only seven configurations owing to the high computational cost: on average, each configuration requires 15,131 CPU core hours.
Further details of the DFT-SSCHA calculation parameters are reported in Supplemental Information Sec.~III.
As discussed in Sec.~\ref{sec:method-df}, the errors arise from two levels of approximation; accordingly, both the error between DP-SSCHA and DFT-SSCHA and the error between DF and DP-SSCHA are reported.
The MAEs in free energy, together with the corresponding forces and virials, are illustrated in Fig.~\ref{fig:fig2}.

The DF (MT) model achieves a test MAE of $14.0$~meV/atom for free energy ($A$), $25.7$~meV/\AA{} for forces ($F$), and $17.9$~meV/atom for virial ($\Xi$) with respect to DFT-SSCHA.
This is a natural consequence of the well-controlled errors at both levels of approximation: the DP-SSCHA approximation to DFT-SSCHA and the DF approximation to DP-SSCHA.
Moreover, the free energy and free-energy force errors are dominated by the discrepancy between DP-SSCHA and DFT-SSCHA, whereas the free-energy virial error is dominated by the discrepancy between DF and DP-SSCHA.
Consequently, the DF-versus-DFT-SSCHA MAEs in free energy and forces are closer to those of DP-SSCHA versus DFT-SSCHA, while the virial MAE is closer to that of DF versus DP-SSCHA.
The DF (ST) model, by contrast, exhibits a 36\% higher free-energy error ($19.1$~meV/atom) and a 78\% higher virial error ($31.9$~meV/atom) relative to DF (MT), while their force errors are comparable ($26.6$ vs $25.7$~meV/\AA{}), consistent with the overfitting observed in Table~\ref{table1}.
The test accuracy of the DF (MT) model indicates that it provides a reliable means of evaluating the thermodynamic stability of crystal structures while accounting for nuclear-quantum and finite-temperature effects.

\subsection{Performance}

\begin{table}
  \begin{threeparttable} %
  \caption{Computational cost of relaxing the \hmn{Fm-3m} \ce{LaH10} structure on the free-energy surface using the DF, DP-SSCHA, and DFT-SSCHA methods.
    DF and DP-SSCHA calculations were performed on a single NVIDIA V100 GPU (4.5~CNY per GPU hour); DFT-SSCHA calculations were performed on a 24-core Intel Xeon 8163 machine at 2.50~GHz (2.56~CNY per machine hour).
    $M_s$ denotes the ensemble size in Eq.~\eqref{eq:sscha-esti}.
    h stands for hours and CNY for Chinese Yuan.
    Relative costs are normalized to the DF model.
  } \label{table3}
  {
  \setlength{\tabcolsep}{4.5 mm}{
  \begin{tabular}{cccc}
  \toprule
  \hline\hline
Method & DF & DP-SSCHA & DFT-SSCHA \\
\hline
Device & GPU	& GPU & CPU\\
$M_s$   & $-$	    & 1024 & 100--200\\
Relaxation steps &6	& 15  & 14 \\
Time/step (h/step) & $2.47\times 10^{-5}$	& $3.37\times 10^{-2}$&  $3.20\times 10^{1}$\\
Total Time (h) & $1.48\times 10^{-4}$	& $5.05\times 10^{-1}$&  $4.48\times 10^{2}$\\
Total Cost (CNY) & $6.68\times 10^{-4}$	& $2.27\times 10^{0}$&  $1.15\times 10^{3}$\\
Relative Cost& $1.00\times 10^{0}$	& $3.40\times 10^{3}$&  $1.72\times 10^{6}$\\
\hline\hline
\bottomrule
  \end{tabular}
}
}
  \end{threeparttable}
\end{table}

The DF model substantially improves the computational efficiency of free energy evaluation, thereby accelerating the relaxation of crystal structures on the FES.
This acceleration is quantitatively assessed in Table~\ref{table3}, which compares the relaxation cost of a representative structure, the \hmn{Fm-3m} \ce{LaH10} configuration, using the DF, DP-SSCHA, and DFT-SSCHA methods.

The DF and DP-SSCHA structure relaxations are both performed on a single NVIDIA V100 GPU, ensuring direct comparability.
Since the DF and DP models share similar architectures, the time required to evaluate a single configuration is closely comparable.
However, each DP-SSCHA relaxation step requires $M_s = 1{,}024$ evaluations of the DP model to estimate the mean potential via Eq.~\eqref{eq:sscha-esti}, whereas the DF model yields the free energy, free-energy forces, and free-energy stresses in a single forward pass.
This difference accounts for the three-order-of-magnitude gap in per-step cost ($3.37\times 10^{-2}$~h vs $2.47\times 10^{-5}$~h).
The DF relaxation also converges in fewer steps (6 vs 15) under the same convergence criterion (step-wise free-energy change below 0.1~meV/atom), which may be attributed to the use of different optimizers: the Atomic Simulation Environment (ASE)~\cite{ase-paper} for DF and the \texttt{sscha-python} package~\cite{monacelli2021stochastic} for DP-SSCHA.

The DFT-SSCHA method employs a smaller ensemble size ($M_s = 100$--$200$) owing to its high computational cost, with all DFT calculations performed using Quantum ESPRESSO~\cite{giannozzi2009quantum} on an Intel Xeon 8163 CPU with 24 cores.
A stage-wise relaxation strategy is adopted, in which the structure is first relaxed with a smaller ensemble and subsequently with $M_s = 200$.
Since the DF/DP-SSCHA and DFT-SSCHA calculations use different hardware, we compare their monetary costs based on cloud resource prices available to us: 4.5~CNY per GPU hour and 2.56~CNY per 24-core CPU machine hour, respectively.
The total cost of relaxing the representative structure is $6.68\times 10^{-4}$, $2.27$, and $1.15\times 10^{3}$~CNY for the DF, DP-SSCHA, and DFT-SSCHA methods, respectively.
DP-SSCHA is approximately $507$ times cheaper than DFT-SSCHA, while the DF method yields a $1.72\times10^6$-fold reduction in cost relative to DFT-SSCHA.
This speedup makes high-throughput CSP directly on the anharmonic free-energy surface practical for the first time.

\subsection{The Predicted Free Energy Convex Hull at 300 K}

\begin{figure}
\begin{center}
\includegraphics[width=1.0\textwidth]{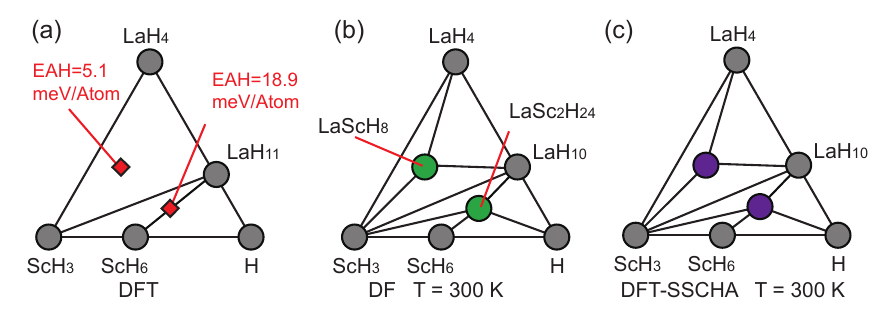}\\[5pt]  
\caption{Convex hulls of the La--Sc--H system at 200~GPa.
(a)~Classical potential energy convex hull calculated by DFT. Red diamonds indicate structures that are metastable on the PES but become stable on the FES, with their energy above hull (EAH) values labeled.
(b)~Free energy convex hull predicted by the DF model at 300~K. Thermodynamically stable structures are marked by green circles. Note that \ce{LaH10} replaces \ce{LaH11} as an on-hull structure compared with the PES hull in (a).
(c)~Free energy convex hull calculated by DFT-SSCHA at 300~K. Thermodynamically stable structures are marked by purple circles.} 
\label{fig:fig3}
\end{center}
\end{figure}

We perform crystal structure prediction at 300~K and 200~GPa across the compositional space La$_m$Sc$_n$H$_x$ using our production-level DF model.
The compositional space of $m \in [1,\, 4]$, $n \in [1,\, 4]$, and $x \in [2(m+n),\, 10(m+n)]$ is randomly sampled. 32,474 independent CALYPSO CSP runs are performed using the DF model as the FES evaluator.
The free-energy convex hull is constructed from all identified FES metastable structures and presented in Fig.~\ref{fig:fig3}(b), where two on-hull structures are identified: \hmn{P6/mmm} \ce{LaSc2H24}, and \hmn{P4/mmm} \ce{LaScH8}.
None of these structures appear on the PES convex hull reported in Fig.~\ref{fig:fig3}(a), which was established in previous work by some of the present authors~\cite{wang2026computational}.
Their stability is confirmed by additional DFT-SSCHA calculations, as reported in Fig.~\ref{fig:fig3}(c).
These findings indicate that nuclear quantum and finite-temperature effects play an essential role in stabilizing \hmn{P6/mmm} \ce{LaSc2H24} and \hmn{P4/mmm} \ce{LaScH8}.

The stability of the \hmn{P6/mmm} \ce{LaSc2H24} structure was previously investigated at 150--300~GPa and zero temperature using DFT-SSCHA calculations~\cite{RN702}.
This phase has since been synthesized experimentally, and its above-room-temperature superconductivity has been verified~\cite{LaSc2H24_syn}.
The DF model successfully predicts \hmn{P6/mmm} \ce{LaSc2H24} as thermodynamically stable on the free-energy convex hull, confirming the reliability of the DF method.

\begin{figure}
\begin{center}
\includegraphics[width=0.6\textwidth]{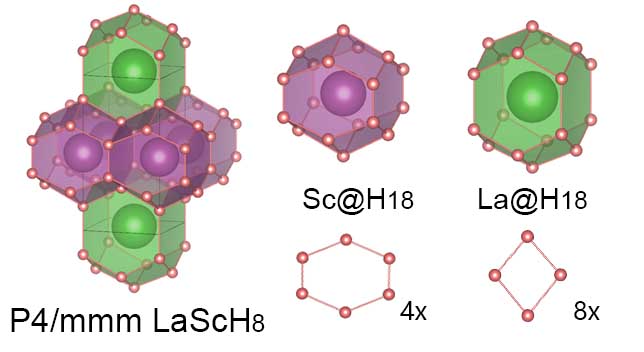}\\[5pt]  
\caption{Crystal structure of the discovered $P4/mmm$ \ce{LaScH8}, in which both La and Sc atoms reside in \ce{H18} cages (green and purple, respectively). Each cage consists of four distorted hexagons and eight rhombi.
} 
\label{fig:fig4}
\end{center}
\end{figure}

The other thermodynamically stable structure on the DF convex hull, \hmn{P4/mmm} \ce{LaScH8}, is also a clathrate hydride (Fig.~\ref{fig:fig4}). It has not been previously reported. 
Tetragonal \ce{LaScH8} adopts the $P4/mmm$ space group, exhibiting lower symmetry than the \hmn{P6/mmm} \ce{LaSc2H24} structure.
The prototype for \ce{LaScH8} was previously identified in \ce{CaScH8} \cite{RN721}.
Lattice information of the phase is presented in Supplemental Information Sec.~IV.
Both La and Sc atoms reside within \ce{H18} cages formed by the same set of hydrogen polygons. 
In \ce{LaScH8}, 5.9\% of the H-H distances are 1.10~Å, while the remaining H-H distances exceed 1.4~Å.


\subsection{Dynamic Stability of Discovered Structures}

\begin{figure}
\begin{center}
\includegraphics[width=1.00\textwidth]{Fig5.pdf}\\[5pt]  
\caption{Free-energy phonon spectra at 300~K, 200~GPa for (a)~\hmn{Fm-3m} \ce{LaH10}, (b)~\hmn{P6/mmm} \ce{LaSc2H24}, (c)~\hmn{P4/mmm} \ce{LaScH8}. Results are shown for DP-SSCHA (black dashed), DF (red solid), and DFT-SSCHA (blue solid).} 
\label{fig:fig5}
\end{center}
\end{figure} 

We calculate the free-energy phonon spectra for the two thermodynamically stable ternary hydrides (\ce{LaSc2H24}, \ce{LaScH8}), and a representative binary hydride (\ce{LaH10}) at 300~K. We compare the results obtained by the DF, DP-SSCHA, and DFT-SSCHA methods in Fig.~\ref{fig:fig5}.
All three configurations, including the newly discovered \ce{LaScH8} phase, exhibit no imaginary frequencies in the spectra predicted by all three methods.
This confirms that these structures are dynamically stable and demonstrates that the DF model accurately captures their dynamical stability at the local minima of the free energy landscape.

The acoustic branches predicted by all three methods show good agreement across the three configurations.
For the \ce{LaScH8} configuration, the DF model yields slightly lower optical-branch frequencies compared to the DP-SSCHA and DFT-SSCHA methods.
This discrepancy may be attributed to noise in the stochastic estimation of the averaged potential via Eq.~\eqref{eq:sscha-esti}, to which the calculation of force constants, as second derivatives of the free energy, is more sensitive.
While this underestimation may affect quantitative predictions of properties sensitive to optical-mode frequencies, such as electron-phonon coupling, it does not compromise the assessment of dynamical stability central to CSP tasks.

\section{Conclusion and Discussion}\label{sec:conclude}

The central finding of this work is that the SCHA free-energy surface, expressed as a function of nuclear centroid positions, shares the same mathematical structure as a potential-energy surface and can therefore be directly learned by a deep neural network potential architecture.
Exploiting this observation, we have constructed the deep free energy model via a two-level concurrent-learning workflow, in which a deep potential model first replaces DFT for SSCHA evaluation, and the resulting SCHA free-energy surface is then learned by a DF model that shares a similar architecture.
A multi-task training scheme further improves the generalizability for accurate free-energy prediction.
Once trained, the DF evaluates free energies, free-energy forces, and free-energy stresses in a single forward pass, entirely bypassing the stochastic sampling of conventional SSCHA and reducing the cost of structural relaxation by approximately six orders of magnitude relative to DFT-SSCHA.

Applied to the La--Sc--H ternary polyhydride system at 200~GPa and 300~K, the DF model achieves test mean absolute errors of $14.0$~meV/atom for free energy, $25.7$~meV/\AA{} for free-energy forces, and $17.9$~meV/atom for free-energy stresses with respect to DFT-SSCHA.
This accuracy and efficiency enable, for the first time, comprehensive CSP directly on the anharmonic free-energy surface across a ternary compositional space.
The DF-based search accurately reproduces the finite-temperature stability of the experimentally observed \hmn{P6/mmm} \ce{LaSc2H24} and discovers another previously unreported stable clathrate hydride: \hmn{P4/mmm} \ce{LaScH8}, with its thermodynamic stability confirmed by independent DFT-SSCHA calculations.
Free-energy phonon spectra computed by the DF model agree well with the DP-SSCHA and DFT-SSCHA results, confirming the dynamical stability of all predicted phases.

Two directions for future development merit discussion.
First, the current framework adopts the SCHA as the underlying free-energy theory, which approximates the nuclear density matrix with a Gaussian ansatz.
While well suited to systems with moderate anharmonicity, this approximation may become insufficient for strongly anharmonic systems where the nuclear density deviates significantly from a Gaussian distribution. 
Second, the DF framework is agnostic to the specific choices of machine-learning potential architecture and CSP algorithm, and can therefore be readily combined with future advances in both areas.

In summary, the DF framework establishes a practical and scalable route for high-throughput crystal structure prediction with nuclear quantum and finite-temperature effects, opening new possibilities for the discovery of materials whose stability is governed by anharmonicity and quantum nuclear fluctuations.

{The code used in manuscript is openly available at github repository \url{https://github.com/Carrotkingdom/DSSCHA.git}.
The DF and DP models are also available at \url{https://www.aissquare.com/models/detail?pageType=models&name=La-Sc-H%3AFESmodels&id=408}.
The datasets are available at \url{https://www.aissquare.com/datasets/detail?pageType=datasets&name=La-Sc-H%3AFESdataset&id=409}.
}

\section{Acknowledgements}
The work is supported by the National Key R $\&$ D Program of China [Grant No. 2022YFA1004300], the National Science Foundation of China [Grant Nos. 12404004, 12561160120, 12525113 and T2225018], and 
the Cross-Disciplinary Key Project of Beijing Natural Science
Foundation No. Z250005.

\bibliography{ref}{}
\bibliographystyle{apsrev4-2}
\end{document}